\documentclass[useAMS,usenatbib]{mnras}
\usepackage{amsmath}
\usepackage{mathtools}
\usepackage{amssymb}
\usepackage{graphicx}
\usepackage{multirow}
\usepackage{color}
\usepackage{longtable}
\usepackage{booktabs}
\usepackage{verbatim} 
\usepackage{multicol,lipsum}
\usepackage{array}
\usepackage{graphicx}
\usepackage{flafter}
\usepackage{placeins}
\usepackage{filecontents}
\usepackage{float}
\graphicspath{ {./images/} }

\title[AGN parameter relations from symbolic regression]{Investigating scaling relations in X-ray reverberating AGN using symbolic regression}

\author[P. Thongkonsing et al.]{P. Thongkonsing$^1$, P. Chainakun$^{1,2}$\thanks{E-mail: \href{mailto:pchainakun@g.sut.ac.th}{pchainakun@g.sut.ac.th}}, T. Worrakitpoonpon$^{1,2}$, A. J. Young$^3$ \\
$^1$School of Physics, Institute of Science, Suranaree University of Technology, Nakhon Ratchasima 30000, Thailand\\
$^2$Centre of Excellence in High Energy Physics and Astrophysics, Suranaree University of Technology, Nakhon Ratchasima 30000, Thailand\\
$^3$HH Wills Physics Laboratory, Tyndall Avenue, Bristol BS8 1TL, UK}

\date{Accepted XXX. Received YYY; in original form ZZZ}

\pubyear{2021}

\begin{document}
\label{firstpage}
\pagerange{\pageref{firstpage}--\pageref{lastpage}}
\maketitle

\begin{abstract}
Symbolic regression (SR) is a regression analysis based on genetic algorithms to search for mathematical expressions that best fit a given data set, by allowing the expressions themselves to mutate. We use the SR to analyze the parameter relations of the X-ray reverberating Active Galactic Nuclei (AGN) where the soft Fe-L lags were observed by {\it XMM-Newton}. Firstly, we revisit the lag-mass scaling relations by using the SR to derive all possible mathematical expressions and test them in terms of accuracy, simplicity and robustness. We find that the correlation between the lags, $\tau$, and the black hole mass, $M_{\rm BH}$, is certain, but the relation should be written in the form of $\log ({\tau}) = \alpha + \beta (\log{(M_{\rm BH}/M_{\odot})})^{\gamma}$, where $1 \lesssim \gamma \lesssim 2$. Moreover, incorporating more parameters such as the reflection fraction ($RF$) and the Eddington ratio ($\lambda_{\rm Edd}$) to the lag-mass scaling relation is made possible by the SR. It reveals that $\alpha$, rather than being a constant, can be $-2.15 + 0.02RF$ or $0.03(RF + \lambda_{\rm Edd})$, with the fine-tuned different $\beta$ and $\gamma$. These further support the relativistic disc-reflection framework in which such functional dependencies can be straightforwardly explained. Furthermore, we derive their host-galaxy mass, $M_{\ast}$, by fitting the spectral energy distribution (SED). We find that the SR model supports a non-linear $M_{\rm BH}$--$M_{\ast}$ relationship, while $\log (M_{\rm BH}/M_{\ast})$ varies between $-5.4$ and $-1.5$, with an average value of $\sim -3.7$. No significant correlation between $M_{\ast}$ and $\lambda_{\rm Edd}$ is confirmed in these samples.

\end{abstract}


\section{Introduction}

The X-ray reverberating active galactic nuclei (AGN) allow us to measure the delay between changes in the X-ray reflection from the accretion disc and the X-ray continuum from the corona, the region of hot plasma surrounding the inner disc. This time delay, known as the reverberation lag, offers information of the light-travel distance between the corona and the disc, hence can be used to study the physical processes as well as the geometry of the innermost region closest to the event horizon of the black hole (BH) \citep[e.g.][]{Fabian2009, Zoghbi2010, Uttley2014, Cackett2021}. The short-timescale variations in the reflection-dominated bands (Fe-L, Fe-K and Compton hump bands) tend to lag behind those in the continuum-dominated bands ($\sim 1-5$~keV band) by tens to hundreds of seconds for these AGNs \citep{Demarco2013, Emmanoulopoulos2014, Kara2016}.
Under the standard lamp-post assumption, the isotropic and point-like corona was constrained to be within $\sim 10$ gravitational radii above the black hole \citep[e.g.][]{Chainakun2016, Epitropakis2016, Caballero2018}.

\cite{Demarco2013} investigated the frequency-dependent Fe-L lags in AGN and found that the amplitude of time lags scale with the BH mass. The similar correlation was also found when calculating the lags using Fe-K bands \citep{Kara2016}. \cite{King2017} suggested that the radio Eddington luminosity scales positively with the disc-corona distance and inversely with the X-ray reflection fraction. \cite{Chainakun2022a} reported an anticorrelation between the fractional excess variance ($F_{\rm var}$) in 2--10~keV band and the BH mass, and suggested that the mass can be accurately predicted using the lags and $F_{\rm var}$. The height of the X-ray corona was found to increase with the source luminosity, in particular for IRAS~13224--3809 \citep{Alston2020, Caballero2020, Chainakun2022b}. \cite{Hancock2022} found that the covering fraction, induced by the motion of non-uniform orbiting clouds, is also inversely correlated with the continuum flux and photon index of the coronal emission.

Furthermore, there is a strong empirical correlation between the BH mass ($M_{\rm BH}$) and the stellar velocity dispersion ($\sigma$) as well as the mass of the host galaxy ($M_{\ast}$), which is believed to arise from the co-evolution of black holes and their host galaxies \citep[e.g.][and references therein]{Kormendy1995, Ferrarese2000, Laor2001, Bluck2011, Reines2015, Shankar2020}. In previous literature, the $M_{\ast}$ may represent the bulge mass ($M_{\rm bulge}$), which, in this case, it is assumed that $M_{\ast} \sim  M_{\rm bulge}$. The AGN and host-galaxy mass relationship is typically given a general form of $M_{\rm BH} \propto M_{\ast}^\beta$ (or $M_{\rm BH} \propto M_{\rm bulge}^\beta$). \cite{Laor2001} suggested that this relation could have a non-linear form with $\beta \sim 1.53$. The scaling relation derived by \cite{Shankar2020} also prefers the non-linear relation, but with $\beta \sim 1.31$. On the other hand, some found this scaling relation is consistent with linearity that has $\beta \sim 1$ \citep{McLure2002,Bettoni2003,Marconi2003}. The exact form of the relation is not well-constrained, and the normalization and coefficient of the $M_{\rm BH}$--$M_{\ast}$ relation can be biased due to the sample-selection effects \citep{Bernardi2007, Shankar2016, Shankar2017}.

In this work, we focus only on the X-ray reverberating AGN samples observed by the {\it XMM-Newton} telescope and previously studied by \cite{Hancock2022}. The corresponding $M_{\ast}$ are derived using the data from the Sloan Digital Sky Survey (SDSS) and the Galaxy Evolution Explorer (GALEX). We then apply the symbolic regression (SR) which is a machine learning method based on genetic algorithms to search for the equation that best fits a given set of data \citep[e.g.][]{Cranmer2020, Udrescu2020, Matsubara2022}, and to identify the complex relationships between the derived parameters. In SR, a mathematical equation is represented as a tree-like structure where the variables and constants are depicted at the leaves and the mathematical operations are located at the nodes. The genetic algorithm works by iteratively generating new equations, assessing how well they match the data, and combining the best-fitting equations to produce new ones. The process continues until an equation that accurately describes the relationship between the input and the output variables is found. During this, the SR evaluates and compromises the model accuracy and complexity so that the obtained equations are not too complex, hence the underlying physical processes can still be interpreted. 

By using the SR, we first investigate the lag-mass scaling relation to test its linearity. The ultimate goal is to use the SR to find the best mathematical expressions (in terms of accuracy, simplicity and robustness) that can explain the relations of the AGN parameters as well as the host-galaxy mass. The AGN data and their parameters used in this study are presented in Section 2. The derivation of the host-galaxy mass for these AGN is explained in Section 3. The SR model and the setting for bootstrapping are described in Section 4. We analyse the obtained equations for the parameter relations in Section 5. The discussion and the conclusion are presented in Section 6 and 7, respectively.

%We may also refer to Williams+2021 for LeMMINGs IV: The X-ray properties of a statistically-complete sample of the nuclei in active and inactive galaxies from the Palomar sample.

\section{AGN Data}
We analyse 20 X-ray reverberating AGN that were available in the {\it XMM-Newton} archives. Their parameters investigated here are listed in Table~\ref{tab_data}. The reflection fraction ($RF$), the Fe-L 0.3--0.8 v.s. 1--4 keV band time-lag ($\tau$), the frequency where the lag is seen ($\nu$), the photon index of the X-ray continuum ($\Gamma$) and the Eddington ratio ($\lambda_{\rm Edd}$) are from the spectral-timing analysis of \cite{Hancock2022}. Note that the reflection fraction ($RF$) is defined as the ratio of coronal photons that hit the disc to those that reach infinity. These samples are low redshift AGN ($0.001 \lesssim z \lesssim 0.08$), except for PG1247+267 that has the redshift $z=2.043$ \citep{Bechtold2002} (For more details of these observations, see Appendix~\ref{sec:ApenA}). 

For 9 sources in our samples, the BH masses estimated by the optical reverberation are available in the public web database \citep{Bentz2015}, where we adopt their values and errors using the mean virial factor of $\langle f \rangle$ = 4.3 \citep{Grier2013}. For IRAS~13224--3809, we adopt the BH mass and respective error from \cite{Alston2020} who investigated short-timescale variations and perform simultaneous lag-frequency spectral fitting, and found that the obtained mass is well consistent with previous analysis using the power spectral density \citep{Alston2019}.

For remaining sources, the BH masses were mostly estimated by the velocity dispersion or the empirical formula for a relation between the size of the broad-line radius and the monochromatic luminosity ($R_{\rm BLR}$--$L$) \citep{Kaspi2000, Bian2003}. However, there is intrinsic scatter in the $R_{\rm BLR}$--$L$ relation that could lead to a significant error in determining the mass. For example, emission line widths of H$\beta$ and Mg II lines are usually used to infer the mass in low redshift AGN, but the intrinsic scatter in the $R_{\rm BLR}$--$L$ relation when using H$\beta$ and Mg II lines could be $\sim 0.2$ dex \citep{Bentz2013} and $\sim 0.36$ dex \citep{Homayouni2020}, respectively. Nevertheless, the errors in the BH masses mainly arise from the unresolved structure of the BLR and the uncertainty in the dynamics of the inside gas. Combining all the uncertainties, the significant error in the mass estimation could rise up to $\sim 0.5$ dex, which is still small in comparison with the intrinsic scatter in the mass of the sample \citep{Krolik2001, Wang2001}. We then use the estimate on the BH mass error of 0.5 dex for these sources. The calculations for the host-galaxy total stellar mass ($M_{\ast}$) presented in the last column in Table~\ref{tab_data} are outlined in the next section.

\section{Host-galaxy mass}
We also expand the data by including the observed parameters of their host galaxy \citep{Gaspari2020} to investigate the relations between the multi-scale parameters. This includes the stellar mass of the galaxy ($M_{\star}$). To estimate $M_{\star}$, we employ the Code Investigating GALaxy Emission (CIGALE) that is publicly available (see \citealt{boquien2019} for details). This code is able to model the far ultraviolet (FUV) to radio galaxy spectral energy distribution (SED), i.e. the SED fitting, from the input photometric data from observations in different filters and estimate many galaxy physical properties, including the stellar mass. The inputs for the SED fitting comprise the ugriz fluxes from the Sloan Digital Sky Survey (SDSS) and, if available, the FUV and NUV fluxes from the Galaxy Evolution Explorer (GALEX). For objects whose SDSS photometric data are missing, we adopt the UBVR fluxes from the SIMBAD database instead. The redshift of the object for the fitting is from the SDSS database or the spectroscopic redshift from SIMBAD database in the case where the redshift is not provided by the former database.

\begin{figure*}
\centering
\includegraphics[trim={0cm 0 0 0},clip,scale=0.65]{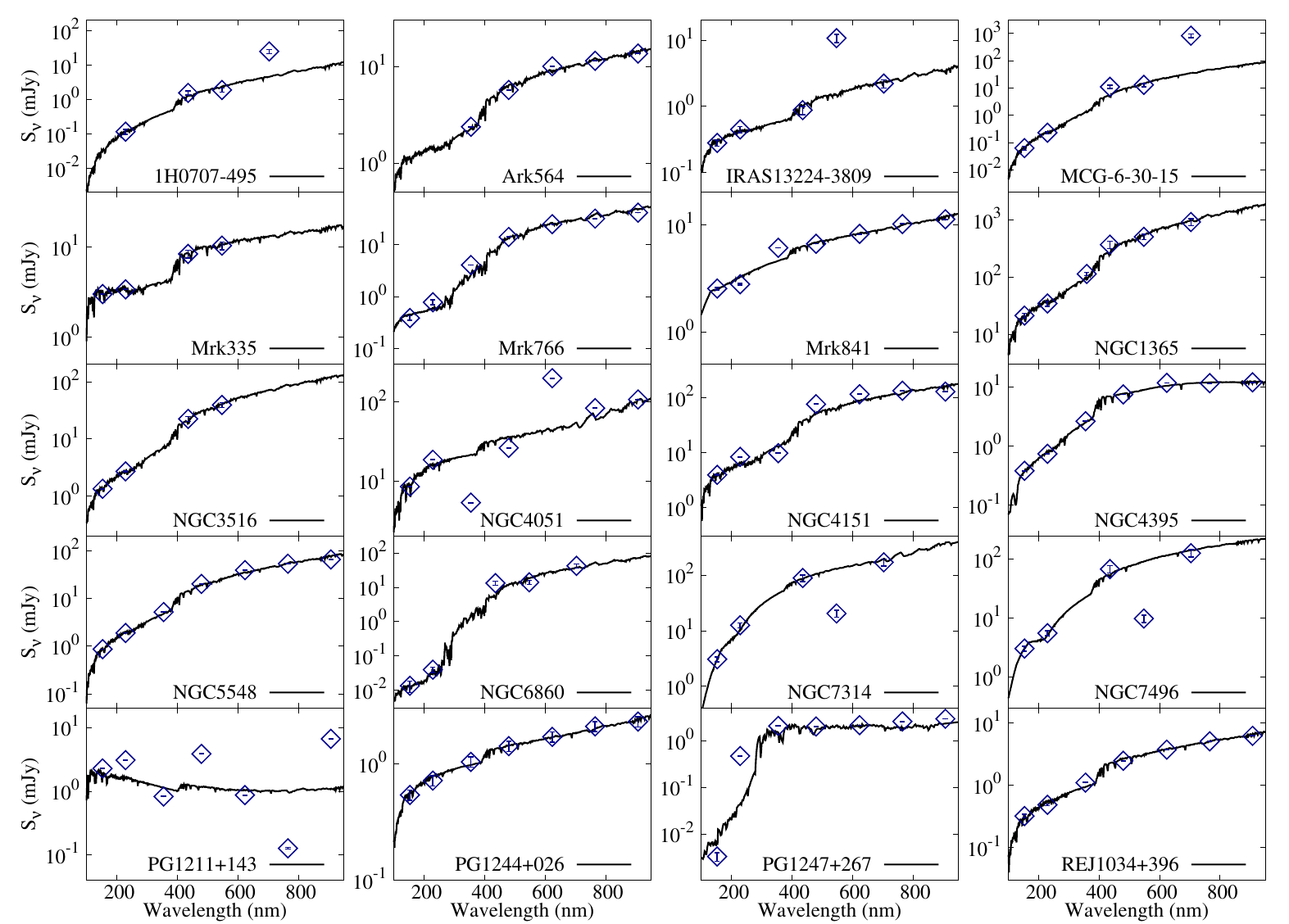}
%\put(-187,160){IRAS 13224-3809} 
%\vspace{-0.55cm}
\caption{Modelled SED in solid line and observed flux densities in points for objects listed in Table.~\ref{tab_data}. Error bars indicate the uncertainty of the observation.
    \label{fig-sed}}
\end{figure*}

In the code, we adopt the exponential star-formation history (SFH) and the stellar population from the libraries of \citet{bc2003} to construct the spectrum of stellar populations. The models for dust attenuation and emission are adopted from \citet{calzetti2000} and \citet{casey2012}, respectively. To model the AGN contribution, we adopt the model of \citet{stalevski2012}.
All modelled SED comparing to the observed flux densities for each object are shown in Fig.~\ref{fig-sed}. The obtained values of the host-galaxy masses are listed in the final column in Table~\ref{tab_data}. We also provide the error of logarithmic $M_{\star}$ that corresponds to the Bayesian-like statistical error estimated from masses of all possible SED models weighted by a factor depending on the chi-square value. We remark two factors that contribute significantly to the errors. The first factor is the number of the input passbands in the way that fewer inputs result in a higher stellar mass error. Secondly, the cases with outlier tend to produce a large error.

Note that the $M_{\star}$ for some of these samples were not measured before. We then verify our code and method for the $M_{\star}$ measurements by using it to determine the stellar mass of some objects in past papers, more specifically some of those in \cite{Reines2015}. In that paper, $M_{\star}$ was estimated using the mass-to-light ratio that depended on the color index. The obtained $M_{\star}$ differs from their values by, on average, $0.28~\rm{dex}$ which is comparable to the offset between $M_{\star}$ estimated from using different color indices.

\begin{table*}
\begin{center}
 \caption{Observed AGN data and associating parameters used in this analysis. The table presents the black hole mass ($M_{\rm BH}$), the reflection fraction ($RF$), the spectral photon index ($\Gamma$), the maximum soft reverberation lag ($\tau$), the frequency where the maximum lag appears ($\nu$), the Eddington ratio ($\lambda_{\rm Edd}$), and the host-galaxy mass ($M_{\ast}$). The numbers in brackets represent the references: (1) \protect\cite{Bian2003}; (2) \protect\cite{Ponti2012}; (3)  \protect\cite{Alston2020}; (4) \protect\cite{Gonzalez2012}; (5) \protect\cite{Schulz1994}; (6) \protect\cite{Marconi2008}; (7) \protect\cite{Lanzuisi2016}; (8) \protect\cite{Done2012}; (R) indicates the optical reverberation mass estimate from \protect\cite{Bentz2015}. } 
 \label{tab_data}
\begin{tabular}{llllllll}
\hline
AGN name & log($M_{\rm BH}/M_{\odot}$) & $RF$ & $\Gamma$ & $\tau$ (s) & $\nu$ (Hz) & $\lambda_{\rm Edd}$ & $\log(M_{\ast}/M_{\odot}$)\\
\hline
1H0707--495 & $6.31 \;\pm \; 0.50$ (1) & $2.14^{+0.15}_{-0.15}$ & $3.38^{+0.025}_{-0.02}$ & $29.1 \pm 3.6$ & $1.55 \times 10^{-3}$ & 1.05 & $10.88 \pm 1.48$ \\

Ark564  & $6.27\; \pm \;0.50$ (2)  & $0.64^{+0.44}_{-0.26}$ & $2.36^{+0.06}_{-0.03}$ & $36.2 \pm 10.5$ & $6.07 \times 10^{-4}$ & 0.976 & $9.91 \pm 0.39$ \\

IRAS13224--3809 & 6.28 $\pm 0.04$ (3) & $3.20^{+0.36}_{-0.37}$ & $3.25^{+0.04}_{-0.02}$ & $39.3 \pm 9.6$ & $5.06 \times 10^{-4}$ & 6.91 & $10.90 \pm 0.65$ \\

MCG--6--30--15 & 6.30$^{+0.16}_{-0.24}$ (R) & $10.00^{+0.00}_{-4.99}$ & $2.00^{+0.02}_{-0.12}$ & $15.9 \pm 5.9$ & $9.66 \times 10^{-4}$ & 0.478 & $10.61 \pm 0.85$ \\

Mrk335	 & 7.23$\; \pm \;0.40$ (R) & $10.00^{+0.00}_{-4.38}$ & $2.82^{+0.28}_{-0.20}$ & $132.7 \pm 36.4$ & $2.65 \times 10^{-4}$ & 0.588 & $10.33 \pm 2.99$ \\

Mrk766	 & 6.82$^{+0.05}_{-0.06}$ (R) & $4.33^{+0.41}_{-0.08}$ & $1.89^{+0.01}_{-0.01}$ & $23.9 \pm 6.7$ & $9.66 \times 10^{-4}$ & 0.233 & $11.28 \pm 0.13$ \\

Mrk841	 & $8.52\; \pm \;0.50$ (2) & $10.00^{+0.00}_{-5.00}$ & $2.00^{+0.49}_{-0.25}$ 
 & $265.9 \pm 217.5$ & $1.02 \times 10^{-4}$ & 0.166 & $10.09 \pm 0.70$ \\ 

NGC1365	 & $7.60\; \pm \;0.50$ (4) & $10.00^{+0.00}_{-6.50}$ & $1.59^{+0.04}_{-0.14}$ & $144.2 \pm 113.4$ & $7.27 \times 10^{-5}$ & 0.0195 & $11.51 \pm 0.37$ \\

NGC3516  & 7.40$^{+0.04}_{-0.06}$ (R) & $8.88^{+1.12}_{-0.54}$ & $1.96^{+0.05}_{-0.07}$ & $256.6 \pm 144.4$ & $7.27 \times 10^{-5}$ & 0.0623 & $10.56 \pm 0.94$ \\

NGC4051	 & 5.89$^{+0.08}_{-0.15}$ (R) & $10.00^{+0.00}_{-4.78}$ & $1.84^{+0.05}_{-0.04}$ 
 & $17.2 \pm 7.1$ & $9.66 \times 10^{-4}$ & 0.0107 & $8.95 \pm 0.16$ \\
 
NGC4151	 & 7.37$ \pm 0.03$ (R) & $10.00^{+0.00}_{-4.53}$ & $1.61^{+0.10}_{-0.12}$ 
 & $488.0 \pm 278.6$ & $1.39 \times 10^{-4}$ & 0.0182 & $12.33 \pm 0.37$ \\
 
NGC4395	 & 5.45$^{+0.13}_{-0.15}$ (R) & $0.40^{+0.00}_{-0.18}$ & $1.06^{+0.00}_{-0.04}$ 
 & $23.9 \pm 16.2$ & $5.06 \times 10^{-4}$ & 0.00423 & $10.15 \pm 0.19$ \\
 
NGC5548	 & 7.72$\pm0.02$ (R) & $6.09^{+3.91}_{-1.29}$ & $1.77^{+0.36}_{-0.36}$ 
 & $156.7 \pm 55.9$ & $2.65 \times 10^{-4}$ & 0.0937 & $11.31 \pm 0.27$ \\
 
NGC6860	 & $7.60\; \pm \;0.50$ (4) & $2.19^{+1.68}_{-0.89}$ & $3.20^{+0.20}_{-0.31}$ & $186.7 \pm 192.5$ & $1.94 \times 10^{-4}$ & 0.0102 & $11.35 \pm 0.31$ \\

NGC7314	 & $6.70\; \pm \;0.50$ (5) & $0.68^{+0.14}_{-0.13}$ & $2.09^{+0.04}_{-0.06}$ & $1.6 \pm 5.8$ & $1.84 \times 10^{-3}$ & 0.0151 & $9.72 \pm 2.77$ \\

NGC7469	 & 6.96$\pm0.05$ (R) & $0.30^{+0.14}_{-0.08}$ & $2.39^{+0.14}_{-0.29}$ 
 & $82.2 \pm 51.1$ & $3.71 \times 10^{-4}$ & 1.11 & $9.73 \pm 1.07$ \\

PG1211+143	 & $7.61\;\pm \; 0.50$ (2) & $2.61^{+2.15}_{-1.18}$ & $2.06^{+0.07}_{-0.07}$  & $215.6 \pm 112.7$ & $8.33 \times 10^{-5}$ & 2.88 & $10.15 \pm 0.27$ \\

PG1244+026	 & $7.26\;\pm \; 0.50$ (6) & $7.61^{+2.39}_{-3.29}$ & $1.94^{+0.33}_{-0.27}$  & $54.5 \pm 20.3$ & $5.06 \times 10^{-4}$ & 0.182 & $9.66 \pm 0.71$\\

PG1247+267	 &  $8.92\;\pm \; 0.50$ (7) & $10.00^{+0.00}_{-6.41}$ & $2.53^{+0.57}_{-0.28}$  & $498.6 \pm 513.2$ & $1.16 \times 10^{-4}$ & 2.09 & $14.37 \pm 0.39$ \\

REJ1034+396	 & $6.18\;\pm \; 0.50$ (8)  & $10.00^{+0.00}_{-4.19}$ & $1.54^{+0.24}_{-0.27}$  &  $55.6 \pm 68.5$ &  $2.65 \times 10^{-4}$&0.660 & $10.42 \pm 0.25$ \\

\hline
\label{table-data}
\end{tabular}
~\\
\end{center}
\end{table*}
\nopagebreak

\section{Symbolic regression and Bootstrapping}
\label{sec:SR}

The symbolic regression (SR) from the {\tt PySR}\footnote{\url{https://github.com/MilesCranmer/PySR}} module \citep{Cranmer2020} is applied to discover and evaluate the relations among the parameters listed in Table~\ref{tab_data}. The mathematical operators and functions are combined to create the symbolic expression. The SR is not limited to certain data types or assumptions regarding variable relationships, so it can be applied to a wide range of research areas. For example, \cite{Delgado2022} studied the relationship between galaxies and haloes by using the SR to incorporate the secondary halo parameter to the standard halo occupation distribution model. \cite{Butter2021} used the SR to assist in discovering an interpretable equation of the parameters from the Large Hadron Collider (LHC) experiment process. \cite{Wang2022} used the SR for feature engineering to extract explicit expressions for the band gap energy.

The input consists of a set of independent variables (features) and a dependent variable (target). Firstly, the SR generates a population of random mathematical equations, which is defined by {\tt population}. The equation is represented as a tree-like structure composed of multiple nodes of mathematical operators and leaves of variables or constants. The operators are further divided into the unary and binary operators. For simplicity, we employ the set of ($+$, $-$, $\times$, $\div$) and ({\tt square}\footnote{{\tt square}$(x)=x^2$}, {\tt sqrt}\footnote{{\tt sqrt}$(x)=\sqrt x$}, {\tt log}\footnote{{\tt log}$(x)=\log_{10}(x)=\log(x)$}, {\tt exp}\footnote{{\tt exp}$(x)=e^{x} \; ; \; e \sim 2.718$}) to be the binary and unary operators, respectively. We also include an additional operator {\tt pow}\footnote{{\tt pow}$(x,y)=x^y$} to test if the parameter relations require a more complex operator. The complexity of the model increases by 1 for each inclusion of one operator, variable or constant. The maximum complexity allowed for the model is controlled by {\tt maxsize}. Here, we fix {\tt population} $=20$ and {\tt maxsize} $=20$.

The equations in the population are used to create new equations through the crossover process in which elements of the equations are merged to generate new, possibly better-performing equations. The newly created equations are subjected to mutation which can be induced by making arbitrary adjustments or random changes to the equations in order to generate fresh, possibly more effective, solutions. The new population of equations is evaluated and the process is repeated until the maximum number of iteration defined by {\tt niterations} is reached. The probability for the crossover process and mutation is set as the default value in the {\tt PySR} module.

In principle, a more complex equation tends to provide a more accurate fit to the data than all simpler equations. Therefore, to avoid overfitting, each equation is evaluated based on the score which compares how loss of the discovered equation reduced with increment in complexity \citep{Cranmer2020}
\begin{align}
{\rm score} = -\frac{\Delta\log{{(\rm MSE})}}{\Delta c} \; ,
\label{eq:score}
\end{align}
where MSE is the mean squared error defined as the average of the square of the best-fit residuals, and $c$ is the complexity depending on the total number of operators, variables and constants in the equation. The score is then the indicator that weighs the improved MSE and the increase in complexity for the newly-generated equation (i.e. the model is jointly optimized for both simplicity and accuracy). Note that the score is dependent on the entire set of the discovered equations. Furthermore, the weight factor ($w$) that regulate the importance between the AGN samples is applied to deal with the uncertainty of the observed parameters (e.g. $w=1/\sigma^2$ where $\sigma$ is the standard deviation, so that the AGN that has smaller $w$ is less important). In this work, the uncertainties are combined if applicable, and the weight factor $w$ already implemented in the SR algorithm is used for the MSE calculation. Nevertheless, the SR is an iterative and stochastic process, meaning that the specific equations and the sequences in which they are formed may vary each time the algorithm is executed. Therefore, the best-performing equations are compared and chosen after running the algorithm numerous times. 

Despite the fact that the search space for mathematical expressions for a particular problem can be very large, traditional methods typically demand prior knowledge or specific assumptions about parameter relationships beforehand. On the other hand, the SR has the advantage that it can derive the relations without first knowing the form of the equation. It is then not limited to specific assumptions about the parameter relationships. More importantly, the SR can still incorporate prior knowledge (e.g., preference of operators, basis functions, or specific variables) if necessary. For example, if we know that there is a physical relation between the BH mass and time lags, we can set preference of the BH mass to be one of the features in explaining the time lags. In fact, we can put a constraint for the time-lag equation to be any given function of mass. Here, we do not put a constraint on this preference since the SR itself should find the most relevant features and relations for the given problem. The other features such as $RF$ and $\Gamma$ can appear in the equation as well if they help improve the fits. The SR then has a potential to reveal alternative relations or even discover the hidden parameters, if they exist. Finding a coherent relationship between these parameters would further support the relativistic disc-corona model, rather than, e.g., absorption or other models in which such a scaling might not be so straightforward to explain.

Since our data contain only 20 AGN samples, we also analyze the results in terms of the robustness by using a bootstrap resampling technique. This works by generating a new data set of the size $n=20$ where the data are randomly picked from the original data set in the way that some original samples can appear once, twice, or do not appear at all in the new data set.
In our analysis, the bootstrapping is performed by using the resample function {\tt sklearn.utils.resample} in the {\tt scikit-learn} module\footnote{\url{https://scikit-learn.org/}} \citep{Pedregosa2012}. The aim is to explore a more unbiased and possibly simpler form of the parameter relation.

\section{Results and analysis}

\subsection{Lag-mass scaling relation}

The data reveal that the Pearson correlation coefficient of the lag and the mass $r_{P} = 0.77$ ($p<0.05$), while the Spearman's rank correlation coefficient $r_S = 0.81$ ($p<0.05$). For comparison, the relation between the lag and the BH mass is derived using both linear regression (LR) and symbolic regression (SR). For LR, we set the mathematical expression in the form of $\log{(\tau) = \alpha + \beta \log{(M_{\rm BH}/M_{\odot})}}$, and evaluate the free parameters $\alpha$ and $\beta$. The linear relation is fitted using the orthogonal distance regression method by using the function {\tt scipy.odr} in the {\tt scipy} module\footnote{\url{https://scipy.org}}. This method minimizes the sum of the squared perpendicular distances between the data points and the regression line which is suitable when there are the measurement errors in both independent and dependent variables. Given $\sigma_y$ and $\sigma_x$ is the errors on the dependence ($y$) and independence ($x$) variables, the chi-square is estimated via \citep{Tellinghuisen2020} 
\begin{align}
\chi^2 = \sum_{i} \frac{(  \alpha + \beta x_i - y_i     )^2}{(\sigma_{y,i}^2 + \beta^2 \sigma_{x,i}^2)} \;.
\label{eq:chi-square}
\end{align}
The obtained equation is
\begin{align}
{\rm LR:} \log{(\tau)} & = -2.07[\pm 0.62] \notag\\ &+ 0.56[\pm 0.09]\;\log{(M_{\rm BH}/M_{\odot})} \;,
\label{eq:linear-lag-mass}
\end{align}
yielding a $\chi^{2}/\rm{d.o.f.} = 27.15/18$. The uncertainty represents the standard deviations of the estimated parameters. The coefficient of determination ($R^2$) of this fit is 0.50. Note that $R^2 = 1 - RSS/TSS$, where $RSS$ and $TSS$ are the sum of squares of residuals and the total sum of squares, respectively. It represents the proportion of variance in the dependent variable that can be explained by the model. The best possible $R^{2}$ is 1. It is clear that the LR model suggests the non-linear scaling relation between the lag and the mass (i.e. considering $\tau \propto M_{\rm BH}^{\beta}$ where $\beta \neq 1$).

\begin{figure}
\centering
\includegraphics[trim={0.1cm 0 0 0},clip,scale=0.4]{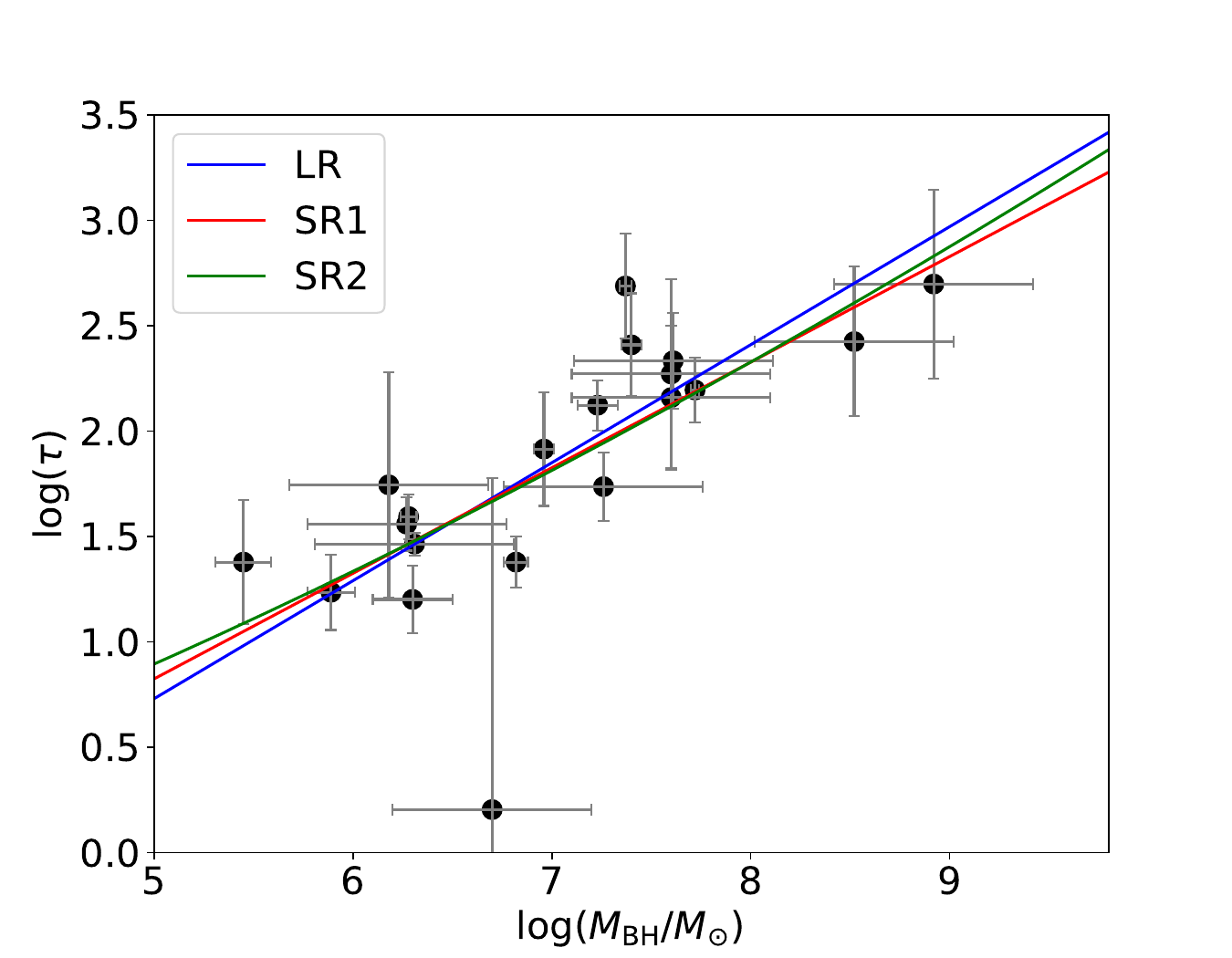}\vspace{-0.45cm}
\caption{Fe-L lag amplitude versus BH mass (black dots). The blue line shows the best fitting LR model (eq.~\ref{eq:linear-lag-mass}), while the red and green lines represent the relations from the SR1 and SR2 models (eqs.~\ref{eq:SR1} and ~\ref{eq:SR2}), respectively. 
\label{fig-lag-sr}}
\end{figure}

For SR, the relation is allowed to be in any forms (with the maximum complexity allowed to be 20, as mentioned in Section~\ref{sec:SR}). While the SR models are evaluated using the score and MSE (eq.~\ref{eq:score}), we also compare their obtained $\chi^{2}$ calculated by taking into account the uncertainty in both dependent and independent variables \citep{Tellinghuisen2020}. We find two SR equations with the complexity of 5 and 6, referred to as SR1 and SR2 models, respectively:
\begin{align}
{\rm SR1:} \log{(\tau)}  & = -1.68 + 0.50 \log{(M_{\rm BH}/M_{\odot})} \;, \;\label{eq:SR1} \\
{\rm SR2:} \log{(\tau)}  & = 
-0.51 + 0.13 (\log{(M_{\rm BH}/M_{\odot})})^{1.5}  \;,
\label{eq:SR2}
\end{align}
with $\chi^{2}/\rm{d.o.f.}=27.49/18$ and $34.98/17$, respectively. The SR1 expression is comparable to the LR equation, suggesting that the SR algorithm can still find the standard equation obtained via the traditional linear regression. The slight differences in the parameter estimates between the LR and SR1 models may arise due to the different minimization methods involved in the fitting processes. However, it suggests an alternative equation, SR2, that can also explain the lags and the mass. The prediction results from the LR, SR1 and SR2 models are illustrated in Fig.~\ref{fig-lag-sr}. Both SR1 and SR2 models have a comparable MSE of $\sim 0.17$. This proves that the SR method can find other models rather than a linear one that can also fit the data reasonably well.

\begin{figure}
\centering
\includegraphics[trim={0.0cm 0 0 0},clip,scale=0.45]{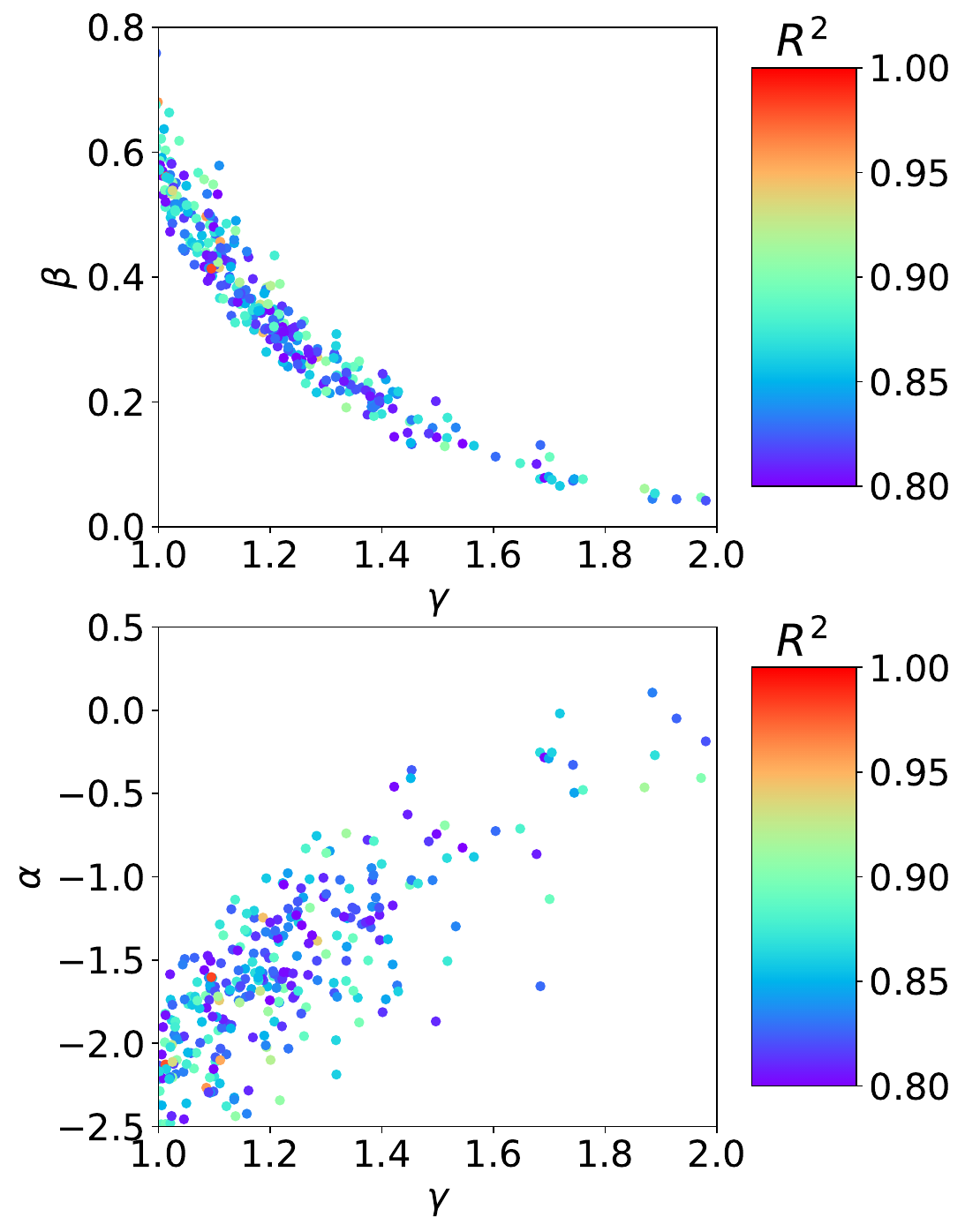}
\put(-188,282){$\log{(\tau)} = \alpha + \beta (\log{(M_{\rm BH}/M_{\odot})})^{\gamma}$} \\
\vspace{-0.1cm}
\caption{Best-fitting $\alpha$, $\beta$ and $\gamma$ from 30,000 bootstrap replicates, when the $\tau$--$M_{\rm BH}$ relation is assumed to be in the form of $\log{(\tau)} = \alpha + \beta (\log{(M_{\rm BH}/M_{\odot})})^{\gamma}$, as suggested by the SR2 equation. The color scale represents the obtained $R^2$.
\label{fig-boot-poly-lag}}
\end{figure}

Fig.~\ref{fig-boot-poly-lag} shows the bootstrapping results for the best fit parameters $\alpha$, $\beta$ and $\gamma$ when we assume their relation is in the form of $\log{\tau} = \alpha + \beta (\log{(M_{\rm BH}/M_{\odot})})^{\gamma}$. Each point represents the best-fitting values from one resampling data set, with the colour scale indicating the accuracy score ($R^{2}$). The result shows how $\alpha$ and $\beta$ change if the $\gamma$ deviates from 1 to 2. Due to a large scatter of the data, bootstrapping suggests that all solutions with $ 1 \lesssim \gamma \lesssim 2$ are still possible ($R^2 > 0.8$). There is no clear difference in the obtained accuracy for particular values of $\alpha$, $\beta$ and $\gamma$. With $\gamma = 1$, the possible $\alpha$ and $\beta$ from bootstrapping cover the values obtained from the LR and SR1 models.

\subsection{Other AGN parameters}
\label{sec:Other AGN parameters}

The intrinsic lag can be suppressed by the contamination between the cross components (continuum and reflection) in both energy bands \citep{Wilkins2013, Kara2014, Chainakun2015}. This is known as a dilution effect that produces a smaller amplitude of the observed lag than that of the intrinsic one. We then investigate the cases when $\tau$ is predicted using more explanatory variables than $M_{\rm BH}$. Interestingly, the obtained equation that contains $M_{\rm BH}$ together with $RF$ and has the lowest loss and complexity can still be written in the form of $\log{(\tau) = \alpha + \beta \log{(M_{\rm BH}/M_{\odot})}}$, with $\alpha = -2.15 + 0.02RF$ and $\beta = 0.55$:
\begin{align}\label{eq:lag-freq-mass}
\log{(\tau)} = -2.15 + 0.02 RF + 0.55 \log{(M_{\rm BH}/M_{\odot})}  \;.
\end{align}
The obtained $\chi^{2}/\rm{d.o.f.}=28.19/17$. This relation can be understood in the way that the reflection fraction controls the amount of the dilution effects \citep[e.g.][]{Wilkins2013}. A larger $RF$ results in a stronger reverberation signal, hence the lag amplitude is more enhanced.

\begin{figure}
\centering

\includegraphics[trim={0cm 0 0 0},clip,scale=0.40]{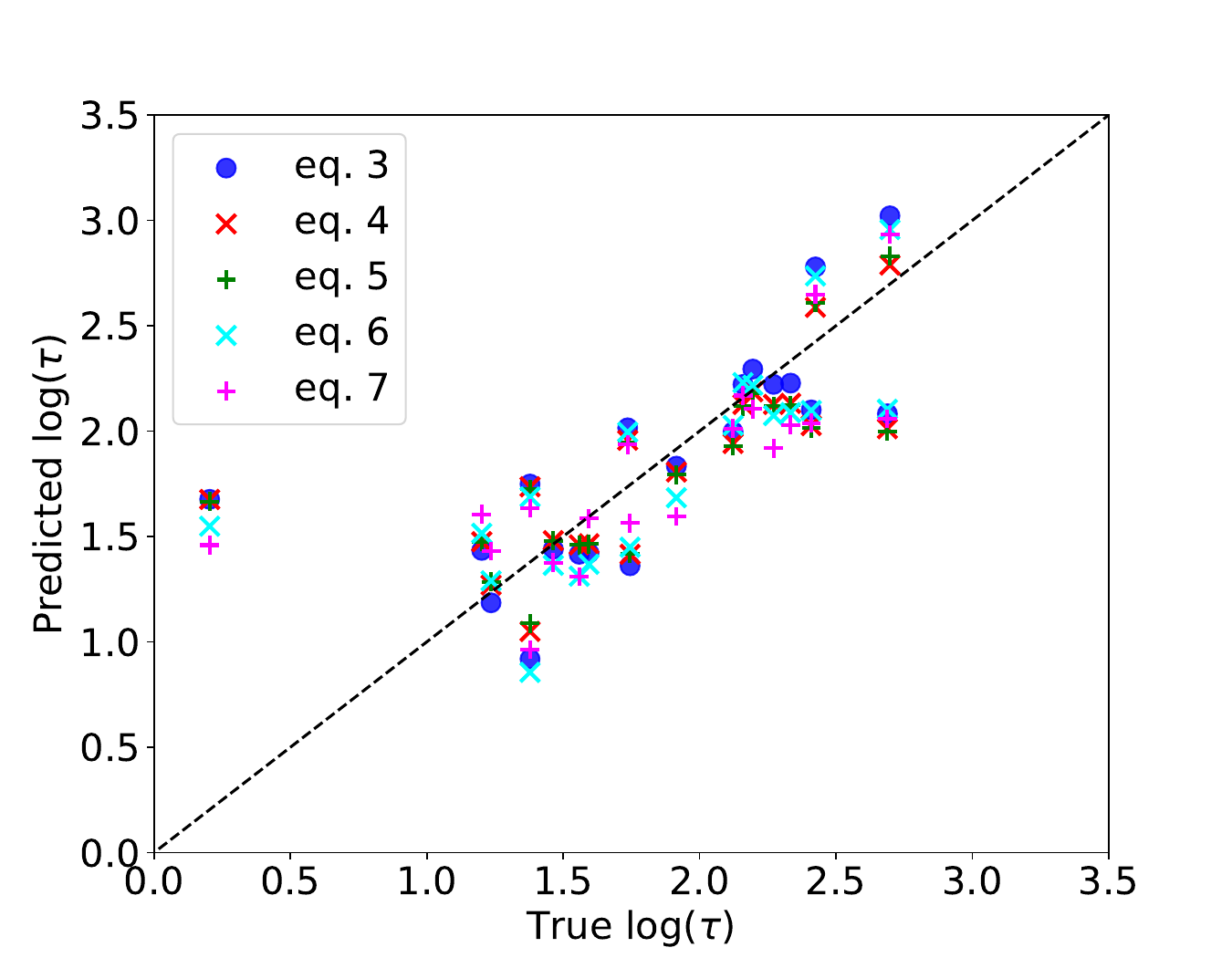}
\vspace{-0.10cm}
\caption{Scatter plots of the true values of the lags versus the predicted time lags obtained from eqs.~\ref{eq:linear-lag-mass}--\ref{eq:final-lag-mass}, showing all model fits are of comparable quality. The black dashed line represents the perfect prediction line. 
    \label{fig-lag-predict-all-comparison}}
\end{figure}

The SR also suggests the expression that involves $M_{\rm BH}$, $RF$ and $\lambda_{\rm Edd}$ to explain the lags. It can be written in the form of $\log{\tau} = \alpha + \beta (\log{(M_{\rm BH}/M_{\odot})})^{\gamma}$, as in eq.~\ref{eq:SR2}, but with the modified coefficients to be $\alpha = 0.03(RF + \lambda_{\rm Edd})$, $\beta = 0.03$ and $\gamma = 2$: 
\begin{align}
\log{(\tau)} = 0.03(RF + \lambda_{\rm Edd}) + 0.03 (\log{(M_{\rm BH}/M_{\odot})})^{2} \;,
\label{eq:final-lag-mass}
\end{align}
yielding $\chi^{2}/\rm{d.o.f.}=51.40/16$. However, the $\Gamma$ does not appear in the suggested SR equations, meaning that after compromising between the accuracy and simplicity, $\Gamma$ may not significantly help improve the fit. This is perhaps because the $\Gamma$ that controls the slope of the X-ray continuum affects mainly on the dilution that is better described by the parameter $RF$. The scatter plots of the actual values of the lags and those predicted by eqs.~\ref{eq:linear-lag-mass}--\ref{eq:final-lag-mass} are shown in Fig.~\ref{fig-lag-predict-all-comparison}. Interestingly, the quality of the fits is good and quite comparable in all models. While the $\chi^{2}$ of eq.~\ref{eq:final-lag-mass} may not that great compared to those from other models, the MSE is found to be $\sim 0.16$, which is comparable to what obtained from eqs.~\ref{eq:SR1} and ~\ref{eq:SR2}. Even though it is not easy to choose which one is the best model, the SR method can show how the reverberation time-lags would depend on not only the BH mass, but also other parameters such as $RF$ and $\lambda_{\rm Edd}$.

%\begin{align}
%\log{(\tau)} = -\log{(\nu)} - 1.511
%\end{align}
% c = 3, loss = 0.044645, score = 0.693798

\subsection{$M_{\rm BH}$--$M_{\ast}$ relation}
\begin{figure}
\centering
\includegraphics[trim={0cm 0 0 0},clip,scale=0.40]{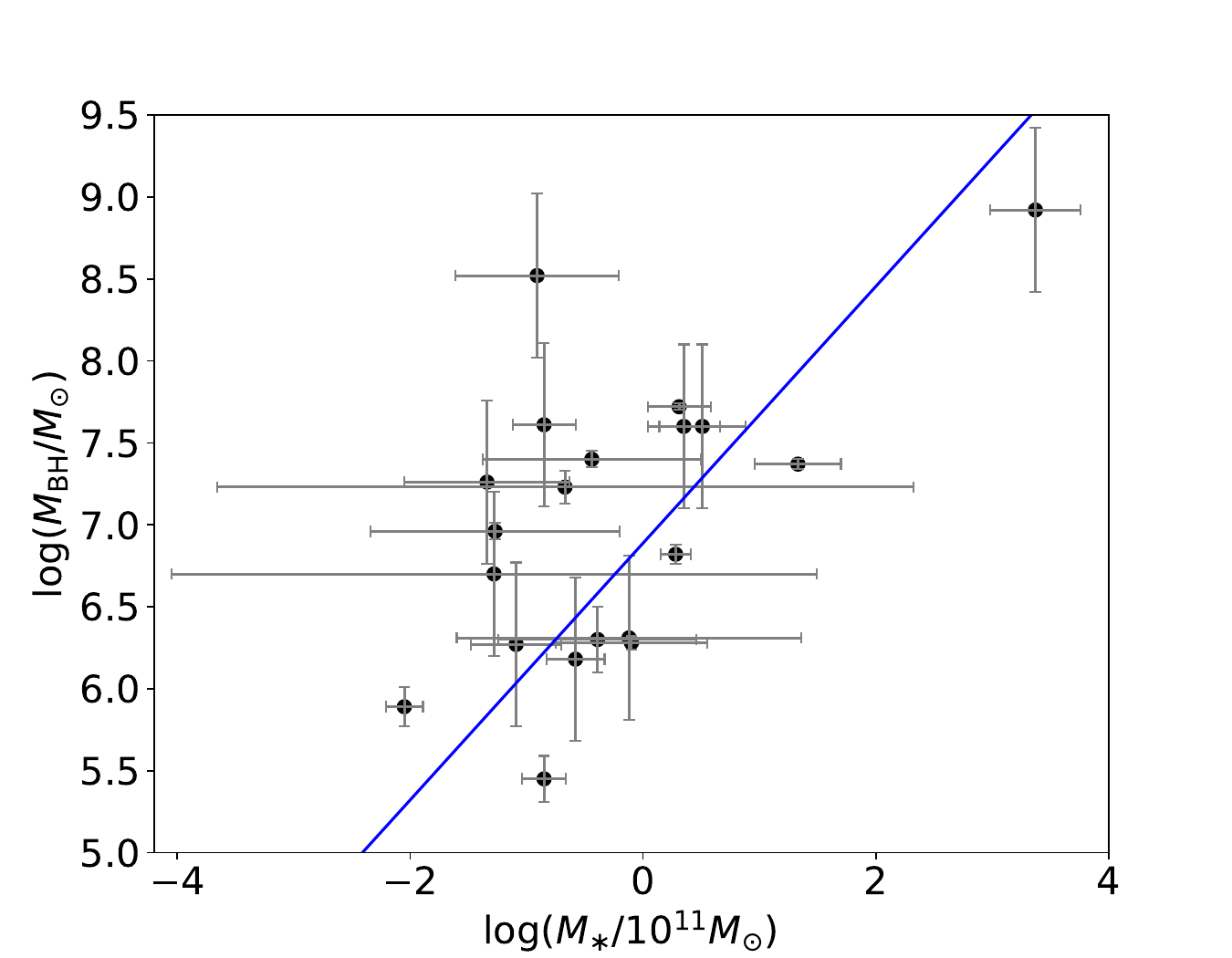}
\put(-75,45){$\alpha = 6.89$} 
\put(-75,35){$\beta = 0.78$} \\
\vspace{-0.10cm}
\caption{Top panel: $M_{\rm BH}$--$M_{\ast}$ relation suggested by the linear regression model (blue solid line) where the black dots represent the AGN samples. The equation is in the form of $\log{(M_{\rm BH}/M_{\odot})} = \alpha + \beta \log{\left(\frac{M_{\ast}}{10^{11}M_{\odot}}\right)}$, where $\alpha = 6.89$ and $\beta = 0.78$. Note that the SR model also prefers the expression in this form. %The uncertainty from the fit is shown in the blue dashed lines.  
    \label{fig-stellar-bootstrap-a}}
\end{figure}

The best relation between the BH mass and the stellar mass of the galaxy obtained by the linear regression is
\begin{align}
\log{(M_{\rm BH}/M_{\odot})} &= \alpha + \beta \log{\left(\frac{M_{\ast}}{10^{11}M_{\odot}}\right)} \;,
\label{eq:stellar}
\end{align}
where
\begin{align}
\alpha = 6.89[\pm 0.15]; \; \beta = 0.78[\pm 0.14] \;,
%\alpha = 7.142; \; \beta = 0.422 \;.
% loss = 0.493210, score = 0.280335, complex = 5
\end{align}
with a $\chi^{2}/\rm{d.o.f.} = 70.97/18$. The fitting result is shown in Fig.~\ref{fig-stellar-bootstrap-a}. The model suggests a non-linear $M_{\rm BH}$--$M_{\ast}$ relation (i.e. $\beta$ is different from 1). In this case, the SR also suggests a relation in the standard form of $M_{\rm BH} \propto M_{\ast}^{\beta}$, similar to when using the linear regression.

\begin{figure}
\centering
\includegraphics[trim={0cm 0 0 0},clip,scale=0.5]{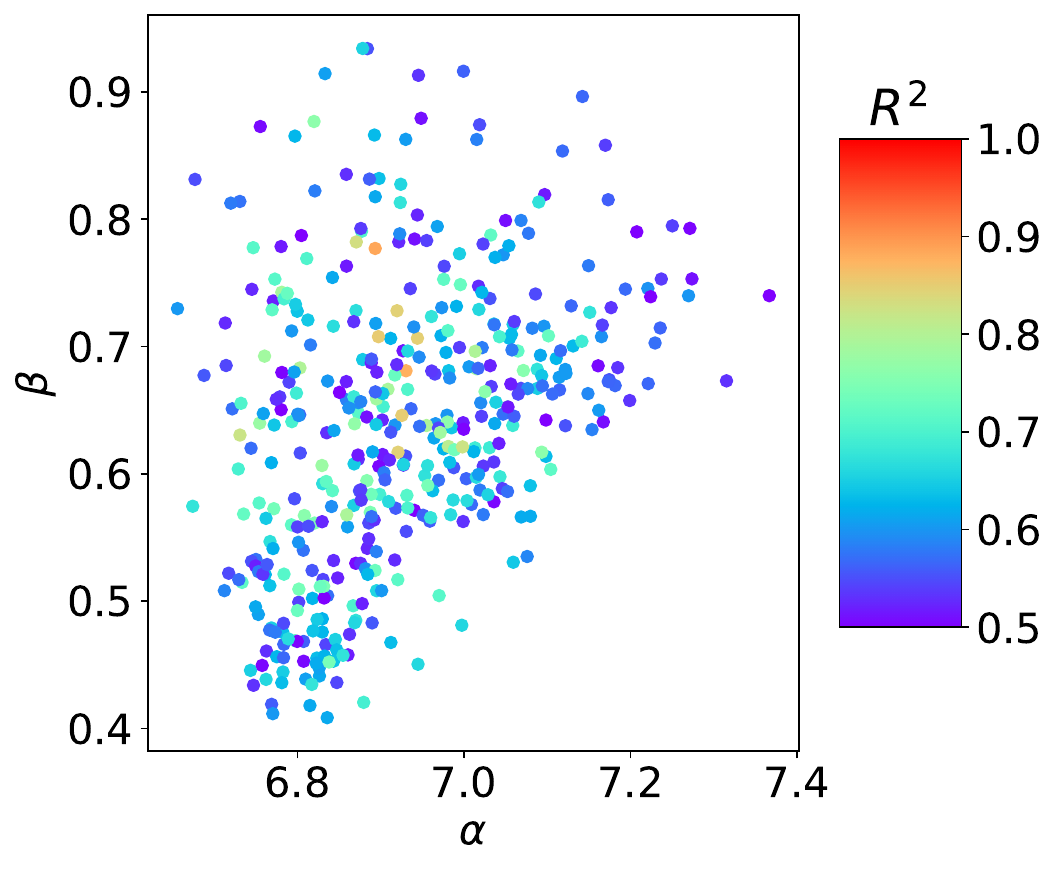}
%\put(-187,160){IRAS 13224-3809} 
\put(-125,195){Bootstrapping,}
\put(-120,185){varying $\alpha, \beta$} \\
\vspace{-0.75cm}
\includegraphics[trim={0cm 0 0 0},clip,scale=0.40]{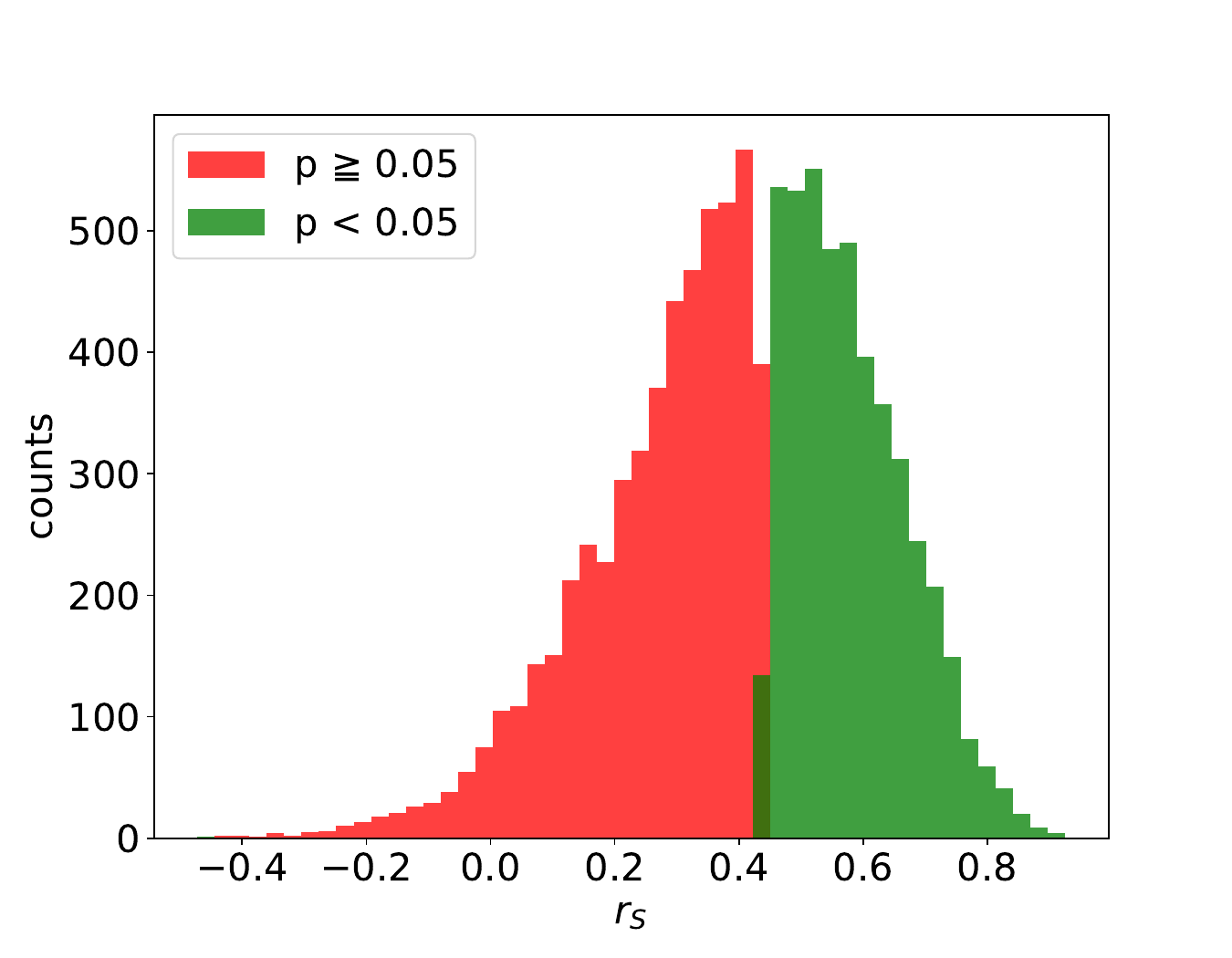}
\put(-220,135){Bootstrapping,}
\put(-220,125){$M_{\rm BH}$--$M_{\ast}$ correlation}
%\vspace{-0.55cm}
\caption{Top panel: Variation of $\alpha$ and $\beta$ based on 30,000 bootstrap replicates when the $M_{\rm BH}$--$M_\ast$ relation is in the form of $\log{(M_{\rm BH}/M_{\odot})} = \alpha + \beta \log{\left(\frac{M_{\ast}}{10^{11}M_{\odot}}\right)}$, with the corresponding $R^2$ shown in the color bar. Bottom panel: The distribution of the obtained $r_{S}$ where the green and red colour represent the significant ($p<0.05$) and insignificant ($p \geq 0.05$) correlation, respectively. 
    \label{fig-stellar-bootstrap-b}}
\end{figure}

Keeping in mind that the number of our AGN samples is small (20 in total), so the bootstrapping is performed 30,000 times to fit 30,000 resampling data sets to see how $\alpha$ and $\beta$ can possibly vary. The result is shown in the top panel of Fig.~\ref{fig-stellar-bootstrap-b}. We find that the obtained equations with $R^{2} > 0.5$ have $\beta$ in the range of $\sim 0.4$--0.9, which still suggest the non-linear $M_{\rm BH}$--$M_{\ast}$ relation, even though the hint of linear solution with $\beta$ reaching $\sim 1$ can be observed. Nevertheless, the monotonic correlation between $M_{\rm BH}$ and $M_{\ast}$ is likely uncertain (Fig.~\ref{fig-stellar-bootstrap-b}, bottom panel). This suggests that the data are too scattered, hence a robust equation may not be easily derived. 

The large scatter in the $M_{\rm BH}$--$M_{\ast}$ relation may arise due to the nature of the sources themselves or the effects of the sample selection. In fact, one source (PG1247+267), located on the most top-right corner in the top panel of Fig.~\ref{fig-stellar-bootstrap-a}, has significantly high redshift of $z=2.043$ compared to other sources, which may lead to a bias in determining the linearity of the $M_{\rm BH}$--$M_{\ast}$ relation. This will be discussed in a later section.

\section{Discussion}

We present an application of the SR to derive the relations of the AGN parameter, and evaluate them in the context of accuracy, simplicity and robustness. Unlike traditional regression approaches that rely on pre-defined mathematical expressions, the SR explores a broad range of possible expressions and functions to find the suitable equations for a given data. \cite{Cranmer2023} showed the capability of the SR in discovering, e.g., the Hubble law, the Kepler law and the Newton law from the measured parameters, illustrating that it can provide the correct relation between the parameters without assuming the forms of the equation in the first place. Above from this, the benefit of the SR is that it can reveal the functional dependencies between variables beyond what we have already known. However, the SR does not provide one most-interpretable solution, but provide a set of solutions ranked with scores. An interpretation of the SR equations still demands expertise and knowledge to reach meaningful conclusions. The best equation might not always be the one with the highest score, but the one that is most accurate and still interpretable.

\cite{Demarco2013} suggested that the $\tau$--$M_{\rm BH}$ relation is compatible with linear scaling if the data with a large mass are treated as upper and lower bounds for the lag amplitude. Our results show that the strong correlation between the amplitude of the lag and the BH mass that has been previously confirmed \citep{Demarco2013, Kara2016, Hancock2022} is quite certain and the non-linear solution is also possible. The BH mass and coronal height increase the lags in a similar way \citep[e.g.][]{Cackett2014, Emmanoulopoulos2014, Caballero2018}. If the corona is fixed at the same gravitational distance above the BH for all these AGN, the lags will be linear scales with the BH mass. The non-linear $\tau$--$M_{\rm BH}$ relation then suggests that the disc-corona geometry varies among the samples.

When adding the other parameters to explain the lags, the fits can be further improved. Interestingly, the SR still suggests the BH-mass based equation in the general form of $\log{(\tau)} = \alpha + \beta (\log{(M_{\rm BH}/M_{\odot})})^{\gamma}$. The $\alpha$ can be either $-2.15 + 0.02RF$ or $0.03(RF + \lambda_{\rm Edd})$, where $\beta$ and $\gamma$ change accordingly. This suggests that the $RF$ that relates to the amount of dilution can act as a correction term that replaces the constant $\alpha$ in the LR equation. With these alternative SR equations, the lags do not only scale with the $M_{\rm BH}$, but also increase with $RF$, which are expected \citep{Uttley2014}. The $\lambda_{\rm Edd}$ can also be an additional factor to modulate the lag-mass scaling relation. In any case, $M_{\rm BH}$ is still the main parameter in these discovered equations.

While a linear model can fit the data well, all model fits (eqs.~\ref{eq:linear-lag-mass}--\ref{eq:final-lag-mass}) are of comparable quality. Therefore, the SR can provide alternative models that explain the data equally well and can establish an independent method to fit the data. It has the ability to reveal the time lags in relation to other parameters beyond the ones we have already known. Recently, \cite{Jaiswal2023} suggested that the effect of increasing the coronal height on the X-ray time delays is quite similar to the effect of the BLR scattering. Nevertheless, our discovered relationship between, e.g., the lags, BH mass and $RF$ supports the reverberation under the disc-corona model since the $RF$ could be tied to the lag-mass scaling relation. The time lags could mainly be produced by relativistic reflection to the disc, rather than, e.g., scattering in the BLR and absorption or reflection to outflowing winds in which such a discovered scaling of $RF$ could be more difficult to explain.

On the other hand, the $M_{\rm BH}$--$M_{\ast}$ relation can be used to study the co-evolution of black holes and their host galaxies, and to improve our understanding of the physical processes that govern the growth and evolution of these objects. The $M_{\rm BH}$--$M_{\ast}$ correlation may be caused by AGN feedbacks (such as jets and winds) that promote star formation and allow BHs to grow at the same rate as their host galaxies, or it may be because the galactic star formation and BH growth both result from the same fuel source. \cite{Ding2020} studied 32 accreting supermassive BHs and their host galaxies at $1.2 < z< 1.7$ by comparing the observational data to the outputs from both hydrodynamic simulation that focuses on AGN feedback and a semi-analytic model that is particularly sensitive to wet galaxy-merger events. They suggested that the scatter of the $M_{\rm BH}$--$M_{\ast}$ correlation could be better explained by the hydrodynamic simulation and that AGN feedback is the result of a causal relationship between the BH and its host galaxy masses. In literature, $M_{\ast}$ may be used as a proxy to the bulge mass ($M_{\ast} \sim M_{\rm bulge}$). The error from this approximation arises if we consider, for example, the Sb/Sbc-type galaxies which show $M_{\rm bulge}/M_{\ast} \sim 0.5$ \citep[e.g.][]{Oohama2009}, as well as the bulgeless galaxies (see also the discussion in \cite{Georgakakis2021}). \cite{Laor2001} investigated the $M_{\rm BH} \propto M_{\rm bulge}^{\beta}$ relation in $\sim 40$ nearby active and inactive galaxies. They found the nonlinear $M_{\rm BH}$--$M_{\rm bulge}$ relation with $\beta = 1.53$, suggesting that $\log (M_{\rm BH}/M_{\rm bulge})$ can increase approximately from $-3.3$ to $-2.3$ with increasing luminosity. Considering that $M_{\ast} \gtrsim M_{\rm bulge}$, the non-linear $M_{\rm BH}$--$M_{\ast}$ relation can also be inferred. Here, by using $M_{\ast}$ rather than $M_{\rm bulge}$, we find no significant correlation between $M_{\ast}$ and $\lambda_{\rm Edd}$ in our samples.

\cite{McLure2002} investigated 72 active galaxies and 18 nearby inactive elliptical galaxies, and found that the BH and the bulge masses are consistent with a linear scaling, with $\beta = 0.95$ and $\log (M_{\rm BH}/M_{\rm bulge}) = -2.9$. \cite{Bettoni2003} used the bulge luminosity and stellar velocity dispersion to infer the central BH masses in nearby radio galaxies, and also reported a strong linearity of the $M_{\rm BH}$--$M_{\rm bulge}$ relation, with $\beta = 0.96$ and the mean $\log (M_{\rm BH}/M_{\rm bulge}) \gtrsim -3.1$. A tight linear correlation between $M_{\rm BH}$ and $M_{\rm bulge}$ for local inactive galaxies was also reported by \cite{Marconi2003}. Moreover, it was suggested that, for the local Universe ($z < 0.1$), $\log (M_{\rm BH}/M_{\ast})$ should be $\sim -3$ \citep{Haring2004, Bluck2011}, assuming $M_{\ast} \sim M_{\rm bulge}$. The samples investigated here are low redshift
AGN ($0.001 \lesssim z \lesssim 0.08$), except PG1247+267 ($z=2.043$). We find that $\log (M_{\rm BH}/M_{\ast})$ varies between $-5.4$ and $-1.5$, with an average value of $\sim -3.7$.

\begin{figure}
\centering
\includegraphics[trim={0cm 0 0 0},clip,scale=0.40]{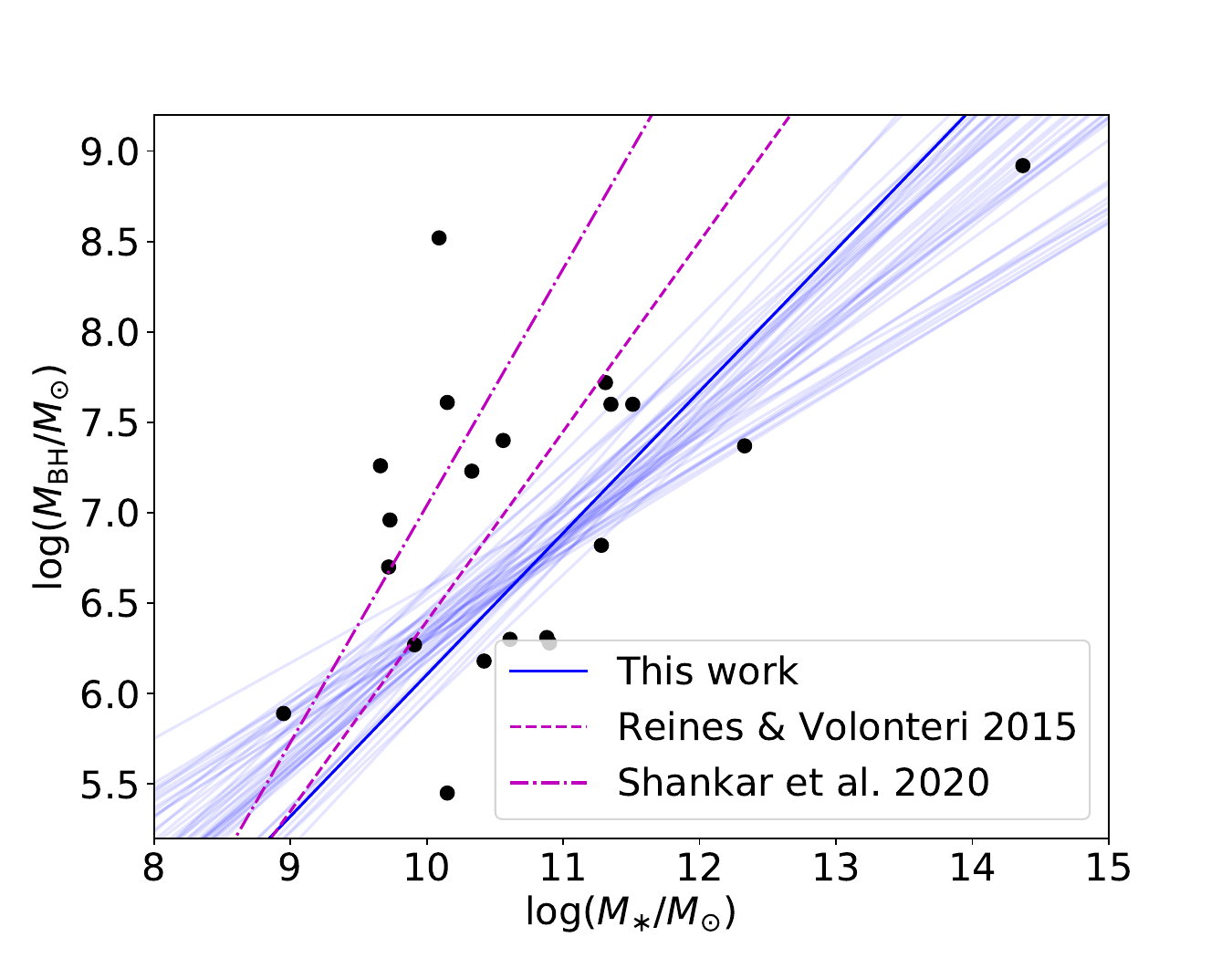}
\includegraphics[trim={0cm 0 0 0},clip,scale=0.40]{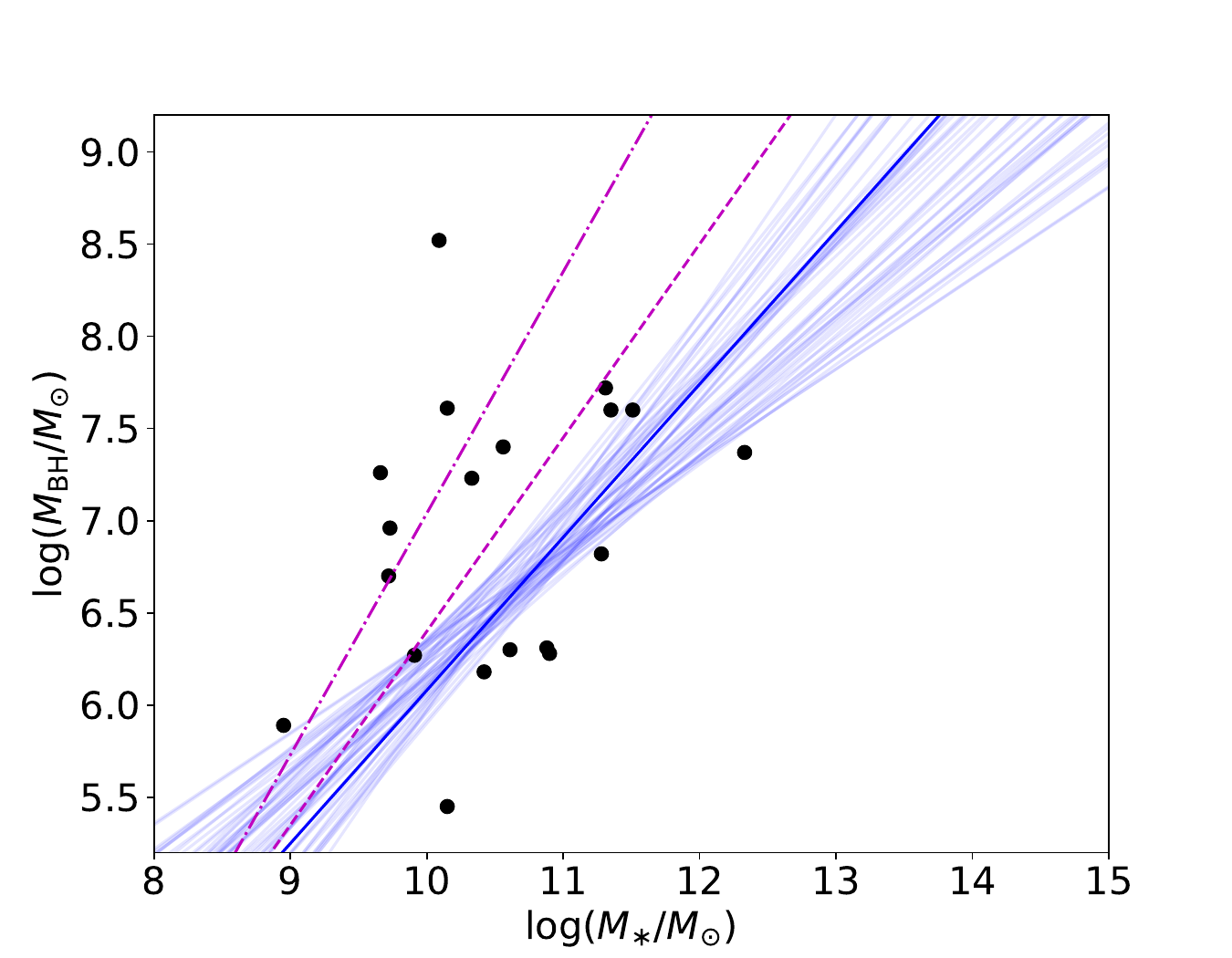}
\caption{Top panel: $M_{\rm BH}$--$M_{\ast}$ relation derived in this work (dark blue line), compared to the results reported in \protect\cite{Reines2015} (magenta, dashed line) and \protect\cite{Shankar2020} (magenta, dashed-dotted line). The light blue lines represent the relations obtained from bootstrap replicates that show $R^2 > 0.5$. The $\beta$ is varied in the range of $\sim0.45-0.80$. The errors are omitted for clarity. Bottom panel: Same as the top panels, but in the case when the sample PG1247+267 (the most top-right data) that has a relatively high redshift of $z=2.043$ is excluded. The $\beta$ is varied in the range of $\sim0.50-0.86$.}
\label{fig-stellar-compare}
\end{figure}

The obtained $M_{\rm BH}$--$M_{\ast}$ relation can be biased due to the selection effects that distort the shape of the scaling profiles \citep{Bernardi2007, Shankar2017}. Here, we compare our $M_{\rm BH}$--$M_{\ast}$ equation with previous studies of \cite{Reines2015} (local AGN samples) and \cite{Shankar2020} (early and late-type galaxy samples). The result is illustrated in the top panel of Fig.~\ref{fig-stellar-compare}. The slope and intercept from our equation is different from previous studies. We also produce 2,000 bootstrap replicates of the data and see how the equations vary among different sample sets. Even in this bootstrapping test, the $M_{\rm BH}$--$M_{\ast}$ relation that is close to that of \cite{Reines2015} cannot be easily obtained. Note that an increase of $M_{\rm BH}/M_{\ast}$ with redshift was also observed \citep[e.g.][]{Peng2006,McLure2006,Merloni2010}. We do not expect the evolution with redshift to be the case here since all samples, excluding PG1247+267, are low redshift AGN. We then investigate the case when PG1247+267 is removed from the data set (Fig.~\ref{fig-stellar-compare}, bottom panel). The resulted equation is slightly closer to the linear relation, and to what was reported in \cite{Reines2015}. Furthermore, we try to fit the data reported in \cite{Reines2015} with the SR method. The obtained results are similar to theirs, suggesting that the use of different methods to fit the data is not the main issue. While it is not clear what drives a non-linear $M_{\rm BH}$--$M_{\ast}$ relation for our samples, it might be that the derived data here are too scattered and might not be representative enough of the AGN population, hence more observed data are still required to draw a robust relation.

We note that the uncertainty in determining the scaling relation may arise from the bias in the $M_{\rm BH}$ and/or $M_{\ast}$ estimation method that is caused by, e.g., the instrumentation limit \citep[e.g.][]{Bernardi2007, Shankar2016, Shankar2017}. Due to the limited information of AGN, the average virial factor was usually used to estimate BH mass but the adopted or derived values may vary among the authors. For example, while we use $\langle f \rangle$ = 4.3 derived from \citep{Grier2013}, \cite{Woo2013} suggested a larger value of $\langle f \rangle$ = 5.1 but \cite{Graham2011} suggested a lower value of $\langle f \rangle$ = 3.8. Note also that the observed flux quality can be another cause of the uncertainty. This can be seen in Fig. \ref{fig-sed} in which in some plots, the discrepancies between some data points and the best SED fit are remarked. More specifically, almost all discrepancies arise from the UBVR manitudes, which are used for the SED fitting for approximately half of the samples, whose flux qualities are mostly classified as the worst according to the SIMBAD database. In addition, the UBVR magnitudes in the SIMBAD database are recompiled from different observations at different times dating from the 80s to recent years. Therefore, the consistency between magnitudes from different filters is questionable. This is not the case for the more recent data sets from GALEX and SDSS that yield a better fit.

Last but not least, the SR can be used to find a relation between BH mass and the lags with the aim to use this relation to measure BH mass. In fact, there are several ways to measure $M_{\rm BH}$ from the X-ray timing data. \cite{Akylas2022} used excess variance to estimate $M_{\rm BH}$ in Seyfert galaxies. They found that the $M_{\rm BH}$ can be calculated from the excess variance if the light curve can satisfy the criteria of S/N ratio and exposure time (i.e. the minimum S/N and duration is $\sim 3$ and $\sim 80$--100~ks). They also suggested to use a light curve in the energy band of 3--10~keV or 10--20~keV, depending on X-ray properties of the source. Meanwhile, \cite{Chainakun2022a} developed a neural network model as an independent method to predict the $M_{\rm BH}$ in the X-ray reverberating AGN. Although the lags in the Fe-K band have been focused, they discovered that the 2--10 keV fractional excess variance and $M_{\rm BH}$ is highly anti-correlated, and the neural network could accurately predict the $M_{\rm BH}$ by using the lag together with the fractional excess variance. Using the lag alone to predict $M_{\rm BH}$, \cite{Chainakun2022a} found that the neural network can still provide a high accuracy of $R^{2} = 0.7302$. The equations for the BH mass prediction from the neural network could not be directly revealed, however. An advantage of using the SR is that it can explicitly reveal the parameter relations in terms of mathematical expressions. Nevertheless, the number of AGN with measured X-ray reverberation lags might not increase significantly in the near future, and there are still other simpler techniques that could be used to predict the mass (i.e. using the width of broad lines or measuring the 2--10 keV excess variance). Despite this, the SR can still establish an independent method to predict the BH mass from the measurement of the X-ray reverberation time-lag as well as to find the relations among the AGN parameters.

\section{Conclusion}

We investigate the scaling relations of the AGN using the SR algorithm. Our data are limited to 20 AGN samples where the soft X-ray reverberation lags are detected. The SR model reveals that the lag-mass scaling relation can be written in the general form of $\log{(\tau)} = \alpha + \beta (\log{(M_{\rm BH}/M_{\odot})})^{\gamma}$. The non-linear relation suggests that the geometry of the corona varies among these AGN samples. Moreover, the SR also reveals some alternative expressions that link more parameters to the lag-mass scaling relation. Rather than being a constant, $\alpha$ can be $-2.15 + 0.02RF$ or $0.03(RF + \lambda_{\rm Edd})$, with accordingly change on $\beta$ and $\gamma$.

The moderate correlation between the host-galaxy mass and the BH mass is observed in this analysis, but it is uncertain. This may lead to difficulty in deriving a robust equation to explain its scaling relation. While our results suggest a non-linear $M_{\rm BH}$--$M_{\ast}$ relation, this can be biased due to the sample selection or due to the small number of the samples. In any case, this work demonstrates a promising approach in using the SR to derive a mathematical expression without first knowing the form of the equation. More data would really help improve the results and place more robust constraint on the true relationships of these parameters.

\section*{Acknowledgements}
This research has received funding support from the NSRF via the Program Management Unit for Human Resources \& Institutional Development, Research and Innovation (grant number B16F640076). P.T. thanks for the financial support from Suranaree University of Technology (SUT) and National Astronomical Research Institute of Thailand (NARIT). P.C. and T.W. thank for the financial support from SUT (grant number 179349).

\section*{Data availability}
The data underlying this article can be accessed from {\it XMM-Newton} Observatory (\url{http://nxsa.esac.esa.int}). The main SR analysis in this work involves the public models/codes adopted from the {\tt PySR} algorithm, available in \url{https://github.com/MilesCranmer/PySR}. The further analysis, including the bootstrapping test, is performed using the modules in {\tt scikit-learn} (\url{https://scikit-learn.org}). Other derived data underlying this article will be shared on reasonable request to the corresponding author.

\bsp
\label{lastpage}

%%%%%%%%%%%%%%%%%%%% REFERENCES %%%%%%%%%%%%%%%%%%

\bibliographystyle{mnras}

\appendix

\section{More observational details}
\label{sec:ApenA}
Full AGN observations used in our analysis and in \cite{Hancock2022} are presented in Table~\ref{table:obs_log}. The first, second and third columns give the name and the redshift of the AGN (1), observation ID (2), and net exposure time of the observations after background subtraction and data screening (3), respectively.
\clearpage
\onecolumn
\begin{longtable}{cccccc}
\caption{The full data including the source, redshift, observation ID and the net exposure time.}\label{table:obs_log}\\
\hline
(1) & (2) & (3) & (1) & (2) & (3) \\
{\textbf{AGN name}} & {\textbf{Obs ID}} & {\textbf{Net exp.}} & {\textbf{AGN name}} & {\textbf{Obs ID}} & {\textbf{Net exp.}}\\
 & & (ks) & & & (ks) \\
\hline
\endfirsthead
\caption* {Table~\ref{table:obs_log}. (Continued.)}
\\
\hline
(1) & (2) & (3) & (1) & (2) & (3) \\
{\textbf{AGN name}} & {\textbf{Obs ID}} & {\textbf{Net exp.}} & {\textbf{AGN name}} & {\textbf{Obs ID}} & {\textbf{Net exp.}}\\
 & & (ks) & & & (ks) \\
\hline
\endhead
\\
1H0707--495 & 0110890201 & 41 & {NGC~4051} & 0109141401 & 106\\
z = 0.0411 & 0148010301 & 76 & z = 0.0023 & 0157560101 & 42 \\
\cite{Zoghbi2010} & 0506200201 & 38 & \cite{Mizumoto2017} & 0606320101 & 45 \\
& 0506200301 & 39  & & 0606320201 & 42 \\
& 0506200401 & 41& & 0606320301 & 21 \\
& 0506200501 & 41 & & 0606320401 & 18 \\
& 0511580101 & 111 & & 0606321301 & 30 \\
& 0511580201 & 93 & & 0606321401 & 35 \\
& 0511580301 & 84 & & 0606321501 & 36 \\
& 0511580401 & 81  & & 0606321601 & 39 \\
& 0653510301 & 112& & 0606321701 & 28 \\
& 0653510401 &  118 & & 0606321801 & 40 \\
& 0653510501 & 93 & & 0606321901 & 36\\
& 0653510601 &  105 & & 0606322001 & 37 \\
& 0554710801 & 86 & & 0606322101 & 24 \\
&& & & 0606322201 & 36 \\
{Ark~564} & 0006810101 & 10 & & 0606322301 & 35 \\

z = 0.024 & 0206400101 & 96 & & &\\ 
\cite{Brandt1994} & 0670130201 & 59 & {NGC~4151} & 0112310101 & 30 \\
& 0670130301 & 55 & z = 0.0033 & 0112830201 & 57\\
& 0670130401 & 55 & \cite{Zoghbi2012} & 0112830501 & 20\\
& 0670130501 & 67 & & 0143500101 & 19 \\
& 0670130601 & 53 & & 0143500201 & 18\\	
& 0670130701 & 41 & & 0143500301 & 19 \\
& 0670130801 & 57 & & 0402660101 & 40\\
& 0670130901 & 56 & & 0402660201 & 34\\
 &&	&& &\\       
{IRAS13224--3809} & 0110890101 & 61 & {NGC~4395} & 0142830101 & 90 \\
z = 0.0406 & 0673580101 & 49 & z= 0.0011 & 0744010101 & 52 \\
\cite{Fabian2013} & 0673580201 & 99 & \cite{Baumgartner2013} & 0744010201 & 48 \\
& 0673580301 & 82 & &&\\
& 0673580401 & 113 & {NGC~5548} & 0089960301 & 84 \\
& 0780560101 & 141 & z = 0.01718 & 0720110801 & 52 \\
& 0780561301 & 127 & \cite{Bentz2009} & 0720110901 & 55 \\
& 0780561401 & 126 & & 0720111001 & 53 \\
& 0780561501 & 126 & & 0720111101 & 35 \\
& 0780561601 & 137 & & 0720111201 & 56 \\
& 0780561701 & 123 & & 0720111301 & 50 \\
& 0792180101 & 123 & & 0720111401 & 52 \\
& 0792180201 & 129 & & 0720111501 & 53 \\
& 0792180301 & 129 & & 0720111601 & 56 \\
& 0792180401 & 120 & &&\\
& 0792180501 & 122 & {NGC~6860} & 0552170301 & 117 \\
& 0792180601 & 122 & z = 0.0149 && \\
&& & \cite{Zoghbi2013} && \\%
{MCG--6--30--15} & 0029740101 & 80 & && \\
z = 0.007749 & 0029740701 & 122 & {NGC~7314} & 0111790101 & 43 \\
\cite{Vaughan2004} & 0029740801 & 124 & z = 0.0048 & 0311190101 & 74 \\
& 0111570101 & 43 & \cite{Schulz1994} & 0725200101 & 122\\
& 0111570201 & 41 & & 0725200301 & 128 \\
& 0693781201 & 121 & &&\\
& 0693781301 & 130 & {NGC~7469} & 0112170101 & 18 \\
& 0693781401 & 49 & z = 0.0164 & 0112170301 & 23 \\
{Mrk~335} & 0306870101 & 120 & \cite{Blustin2003} & 0207090101 & 85 \\
z = 0.0285 & 0600540501 & 80 & & 0207090201 & 78 \\
\cite{Chainakun2015} & 0600540601 & 107 & && \\
&& & &&\\\hline \newpage \\

{Mrk~766} & 0096020101 & 27 & {PG1211+143} & 0112610101 & 53 \\
z = 0.01293 & 0109141301 & 104 & z = 0.0809 & 0208020101 & 46 \\  
\cite{Bentz2009} & 0304030101 & 78 & \cite{Pounds2006} & 0502050101 & 45 \\
& 0304030301 & 98 & & 0502050201 & 35 \\
& 0304030401 & 92 & & 0745110101 & 78 \\
& 0304030501 & 73 & & 0745110201 & 98 \\
& 0304030601 & 85 & & 0745110301 & 54 \\
& 0304030701 & 29 & & 0745110401 & 91 \\
&& & & 0745110501 & 55 \\
{Mrk~841} & 0070740101 & 108 & & 0745110601 & 92 \\
z = 0.0365 & 0070740301 & 122 & & 0745110701 & 96 \\
\cite{Cerruti2011} & 0205340201 & 43 & &&\\
& 0205340401 & 18 & {PG1244+026} & 0675320101 & 123 \\
&& & z = 0.0482 & 0744440201 & 92 \\
{NGC~1365} & 0151370101 & 13 & \cite{Chainakun2017} & 0744440301 & 121 \\
z = 0.0045 & 0151370201 & 2 & & 0744440401 & 127 \\
\cite{Gonzalez2012} & 0151370701 & 8 & & 0744440501 & 118 \\
& 0205590301 & 48 & &&\\
& 0205590401 & 33 & {PG1247+267} & 0143150201 & 32 \\
& 0505140201 & 38 & z = 2.043 &&\\
& 0505140401 & 107 & \cite{Bechtold2002} && \\
& 0505140501(1) & 88 & && \\
& 0505140501(2) & 53 & {REJ1034+396} & 0506440101 & 84 \\
& 0692840201 & 101 & z = 0.04 & 0561580201 & 54 \\
& 0692840301 &  44 & \cite{Done2012} & 0655310101 & 45 \\
& 0692840401  & 87 & & 0655310201 & 50 \\
& 0692840501(1) &  64 & & &\\
 & 0692840501(2) & 34 & & &\\
&& & & &\\
{NGC~3516} & 0107460601 & 114 & & &\\
z = 0.00886 & 0107460701 & 121 & & &\\
\cite{Keel1996} & 0401210401 & 51 & & &\\
& 0401210501 & 61 & & &\\
& 0401210601 & 62 & & &\\
& 0401211001 & 58 & & &\\
&& & & &\\
\hline
\end{longtable}

\twocolumn

\bsp	% typesetting comment
\label{lastpage}
\end{document}